\documentclass[12pt]{article}

\usepackage{hyperref}
\usepackage{bbm}
\def\be{\begin{equation}}
\def\ee{\end{equation}}
\def\ba{\begin{array}}
\def\ea{\end{array}}
\def\bea{\begin{eqnarray}}
\def\eea{\end{eqnarray}}
\def\ap{\alpha^\prime}

\def\ie{{\it i.e.}~}
\def\eg{{\it e.g.}~}
\def\identmat{\mathbbm{1}}

\begin{document}

\thispagestyle{empty}

\begin{center}

{\small\ttfamily ROM2F/2005/02\hspace*{0.8cm} }

\end{center}

%\bigskip

\begin{center}

\renewcommand{\thefootnote}{\fnsymbol{footnote}}

{\bf\Large The open story of the magnetic fluxes}
%\bigskip\bigskip

\textbf{M.~Bianchi} and \textbf{E.~Trevigne}

\vspace{.3cm}

\textit{Dipartimento di Fisica and Sezione INFN \\
Universit\`a di Roma ``Tor Vergata''\\
00133 Rome, Italy}\\
{\texttt{Massimo.Bianchi,Elisa.Trevigne@roma2.infn.it}}
\setcounter{footnote}{0}

\end{center}

\begin{abstract}

We discuss the effects of oblique internal magnetic fields on the
spectrum of type I superstrings compactified on tori. In
particular we derive general formulae for the magnetic shifts and
multiplicities of open strings connecting D9-branes with arbitrary
magnetic fluxes. We discuss the flux induced potential and offer
an interpretation of the stabilization of R-R moduli associated to
deformations of the complex structure of ${\bf T}^6$ in terms of
non-derivative mixings with NS-NS moduli. Finally we briefly
comment on how to extract other low energy couplings and
generalize our results to toroidal orbifolds and other
configurations governed by rational conformal field theories on
the worldsheet.

\end{abstract}

\def\baselinestretch{1.1}

\section{Introduction}
\label{intro}

Constant electromagnetic fields coupled to the ends of open
strings lead to many interesting effects that can be treated
exactly in string theory \cite{ft,acny,nestmag,bp,bachsusy} and
have far reaching consequences in phenomenologically viable vacuum
configurations \cite{dudrev,phen1,phen2,phen3,phen4,phen5,phen6}. The
presence of constant abelian magnetic fields changes the open
string boundary conditions in such a way that the oscillator modes
of charged strings get shifted and the zero modes of neutral
strings (KK momenta and windings) get reshuffled
\cite{asrev,mlgp,lars}.

So far most of the analyses focussed on ``parallel" magnetic
fields $H_a$ on ``factorizable" tori ${\bf T}^d={\bf T}^2 \times
...{\bf T}^2$ or toroidal orbifolds that can be T-dualized to a
single kind of branes intersecting at angles, for reviews see
\cite{kirirev,blumrev,kokorev,uranrev}. The boundary conditions
are coded in the mutually commuting $d\times d$ orthogonal
matrices
$$R_a = (\identmat -H_a)(\identmat +H_a)^{-1} \quad .$$ However, as shown in
\cite{am} for a toroidal compactification of type I strings to
$D=4$, properly turning on oblique but yet abelian magnetic
fields, such that $[R_a, R_b] \neq 0$, in ``non-factorizable"
subtori with common directions may lead to closed string moduli
stabilization while keeping some residual supersymmetry. These
configurations are T-dual to bound states of several different
kinds of intersecting D-branes \cite{raba,mmms,braninbran}.

Here, we present a detailed analysis of the resulting open string
spectrum and discuss the resulting potential for the closed string
moduli fields, as well as their mixing and stabilization. In order
to fix our notation and describe some of the relevant issues in a
simpler setting we start by reviewing the very familiar case of a
2-dimensional torus in section \ref{tdue} and the less familiar
case of a 3-dimensional torus, where oblique magnetic fields first
show up, in section \ref{ttre}. In both these cases, however,
turning on magnetic fields breaks all supersymmetries and prevents
the existence of any stable configuration at finite internal
volume even if one introduces exotic $\Omega$-planes. We discuss
the more interesting case of a 4-dimensional torus in section
\ref{tquat} where we derive a very general formula for the mode
shifts in the presence of arbitrary magnetic fields and argue that all
closed string moduli, except the overall volume and the dilaton,
can be stabilized in a supersymmetric fashion. Scale invariance of
the (anti) self-duality conditions should be held responsible for
the arbitrariness of the overall volume. In section \ref{tcinq},
we sketch how to embed these neat results into a 5-dimensional
torus. All this is meant to be propaedeutic to the analysis of the
phenomenologically relevant case of a 6-dimensional torus studied
in section \ref{tsei}. After some general considerations, we
restrict our attention to supersymmetric configurations and
describe how to compute in full generality the magnetic shifts for
oblique yet supersymmetry preserving magnetic fields. We also
clarify the mechanism of stabilization of R-R moduli associated to
deformations of the complex structure in terms of non-derivative
mixings with NS-NS moduli. To this end we adapt and generalize the
closed string scattering amplitudes on D-branes computed in
\cite{gm,hk} to our case and derive similar formulae for
scattering on $\Omega$-planes \cite{kls}.  We present our
conclusions and perspectives for future investigation in section
\ref{conclus}. In particular, we sketch how to extract tree-level
and one-loop gauge couplings from (simple modifications of) the
results in the previous sections.  We also discuss some
generalizations to toroidal orbifolds and to vacuum configurations
described by rational conformal field theories on the worldsheet.
Very much as in \cite{mbys}, we argue that magnetic fluxes lead to
twisted representations of the relevant chiral algebra.

{}{}

\section{Two-dimensional magnetized torus, yet again}
\label{tdue}

A magnetic field on ${\bf T}^2$ is necessarily proportional to the
volume (K\"ahler) form $J=dx^1 dx^2 = d\bar{z} dz /2i$. In complex
coordinates, $F= F_{12} dx^1 dx^2 =  i f dz d\bar{z}/2$, and the
corresponding magnetic rotation matrix reads \be R =
(\identmat-F)(\identmat+F)^{-1} = {\rm diag}(e^{2i\beta},
e^{-2i\beta}) \ee where \be \tan(\beta) = f \qquad {\rm so \ that}
\qquad \tan(2 \beta) = {2f\over 1-f^2} \quad .\ee In a real basis,
one would have instead \be
R = \left( \begin{array}{cc} {1-f^2\over 1+f^2} & -{2f\over 1+f^2} \\
{2f\over 1+f^2} & {1-f^2\over 1+f^2} \end{array} \right)  \quad .
\ee

Consistency at the quantum level requires Dirac flux quantization
\be f = {m\over n} {\ap\over r_1 r_2} \ee where  the magnetic
monopole number ($m$) and the ``wrapping'' number ($n$) are
integers, in the absence of a (quantized) NS-NS antisymmetric
tensor background \cite{bpstor,wittor,angtor}. Otherwise $m$ gets
shifted by $B n$ \cite{asrev,mlgp,lars}.

Thanks to the abelian nature of the rotation group in two
dimensions, when several abelian magnetic fields are present the
magnetic shifts of the modes of open strings connecting the stacks
$a$ (with flux $f_a$) and $b$ (with flux $f_b$) are given by
\cite{acny,Marchesano:2004yq,Marchesano:2004xz,dt,bp,nestmag} \be \epsilon_{ab} = {\beta_{ab}\over
\pi} = {1\over \pi} [q_a \arctan(f_a) + q_b \arctan(f_b)]\ee where
$q_a, q_b = \pm 1$ take care of possible images under orientation
reversal, \ie worldsheet parity $\Omega$ \cite{asrev}. Let us
stress that the magnetic rotation \be R_{ab} =
R^{q_a}(f_a)R^{q_b}(f_b) = R (q_a f_a+q_bf_b) = R(2\beta_{ab}) \ee
appearing in the boundary state \cite{divelic}, acts by {\it
twice} the amount determining the magnetic shift,
$\beta_{ab}=\pi\epsilon_{ab}$.

In order to properly count the multiplicities of charged strings,
one has to rescale $N_a$ to $\hat{N}_a = {N}_a/n_a$ as a
consequence of the non-trivial wrapping. In general, the target
space coordinates $X^i$ do not necessarily coincide with the
world-volume coordinates $\sigma^\alpha$ but are rather linear
functions of the latter. The jacobian matrix $W_\alpha{}^i =
\partial X^i / \partial \sigma^\alpha$ of the linear
transformation mapping the world-volume $\Sigma^d$ onto the target
space ${\bf T}^d$ must have integer entries. As a result, $W=
\det(W_\alpha{}^i)=n$, is an integer, \eg $W=1$ for the trivial
wrapping $X^i = \delta^i_\alpha \sigma^\alpha$. The induced
world-volume metric is thus $ {\cal G}_{\alpha \beta} =
W_\alpha{}^i W_\beta{}^j G_{ij}$. Clearly there is a separate
wrapping matrix per each stack of (magnetized) branes
$W_\alpha{}^i_{(a)}$ labeled by the index $a$ and $n_a=\det(W_a)$.

Open strings with $q_b = -q_a$ and $a\neq b$, belong to the
representation $(\hat{N}_a, \hat{N}^*_b)$.  For $q_a = q_b$, which
is possible only in the unoriented case, one gets $(\hat{N}_a,
\hat{N}_b)$.  In both cases the degeneracy of the Landau levels
\be I_{ab} = q_a m_a n_b + q_b m_b n_a \ee coincides with the
first Chern number of the abelian gauge bundle. In the superstring
case, $I_{ab}$ counts the number of charged massless fermions, but
the spectrum contains open string tachyons whenever $f_a\neq 0$
for some $a$.

Neutral strings starting and ending on the same stack $a=b$ of
branes and thus with $q_b= -q_a$ have integer modes but carry
``rescaled" KK momenta $p_i=k_i/n r_i\sqrt{1+f^2}$ with $i=1,2$.
In the unoriented case, doubly charged strings, that start on a
stack $a$ and end on its image $b=\tilde{a}$ under $\Omega$ with
$q_{\tilde{a}} = q_a$ are counted by the first Chern number of the
(anti) symmetric tensor gauge bundle.

T-duality along the $y=x^2$ axis maps D2-branes filling ${\bf
T}^2$ into D1-branes wrapped around a direction forming an angle
$\beta = \arctan( m \widetilde{R}_2/ n R_1)$ with the $x=x^1$ axis
\cite{bdl}. The magnetic shifts (up to a factor of $\pi$) are
T-dual to the intersection angles. In the T-dual picture, $I_{ab}$
counts the number of ``algebraic" intersections \cite{uranrev},
while the magnetically rescaled momenta map to momenta along the
non-trivially wrapped D1-branes and windings in the orthogonal
direction \cite{blumrev,mlgp}.

Very much as for the heterotic string on twisted tori \cite{km},
internal magnetic fluxes can generate a potential for the closed
string NS-NS and R-R moduli, $G_{ij}$ and $C_{ij}$, as well as for
the dilaton $\phi$ and the open string Wilson line moduli $A^a_i$
\cite{kst,3formflux,aftda}. The crucial difference is that for
(unoriented) open strings the potential appears at half closed
string loop order. The Born-Infeld contribution of the D-branes
(disk) and the (negative) tension contribution of the
$\Omega$-plane (projective plane) \cite{leigh,mss} combine to \be
V(\phi, G_{ij}; m_a, n_a) = \tau e^{-\phi} \left( 2\sum_a N_a
\sqrt{\det({\cal G}_a + {\cal F}_a}) \mp 32 \sqrt{\det(G)} \right)
\quad , \ee where ${\cal G}^{a}_{\alpha\beta}$, as discussed
above, is the induced metric on the $a^{th}$ stack of branes,
${\cal F}^a_{\alpha\beta}$ is the world-volume magnetic field \be
{\cal F}^a_{\alpha\beta} =
\partial_{\alpha}A^a_{\beta} -\partial_{\beta}A^a_{\alpha}=
W^{(a)}_\alpha{}^i W^{(a)}_\beta{}^j F^a_{ij} \quad ,\ee and
$\tau$ is the constant (moduli independent) part of the brane
tension. The factor of $2$ takes care of image branes and we have
allowed for exotic $\Omega$-planes with positive tension.
Neglecting open string moduli and the WZ coupling to the R-R
moduli,  $V$ turns out to depend only on the positive-definite
overall volume $\omega = \sqrt{\det(G)}$ and the dilaton $\phi$
but not on the complex structure moduli. Independently of the sign
of the $\Omega$-plane tension, \be V(\phi, \omega; m_a, n_a) =
\tau e^{-\phi} \left( 2\sum_a N_a \sqrt{n_a^2 \omega^2 + m^2_a}
\mp 32 |\omega|\right) \ee displays a runaway behavior to
infinity for $\phi$. $\omega$ runs to infinite, respectively zero,
value for standard, respectively, exotic (anti) $\Omega$-planes.
The R-R tadpole condition, $2\sum_a n_a N_a = \pm 32$, can be
satisfied for standard $\Omega$-planes and exotic
$\bar\Omega$-planes \cite{wittor,dms}.

We will not dwelve  any further on the issue of moduli
stabilization on ${\bf T}^2$ since magnetic fluxes break
supersymmetry anyway and generate tachyons  in the open
superstring spectrum. In passing, the situation does not improve
very much for compactifications on product tori $\times_i{\bf
T}^2_{(i)}$ and orbifolds thereof, which have been the main
subject of investigation so far. The above formulae for the
magnetic shifts, multiplicities and zero-modes can be
straightforwardly generalized. Same for the Born-Infeld and WZ
couplings \cite{leigh,mss}. However the only NS-NS moduli that
appear in the potential are the volume moduli $\omega_i$ and they
cannot be stabilized.

{} {}{}
{} {}
{{}}
{{{}}}

\section{Three-dimensional magnetized torus}
\label{ttre}

In three dimensions a magnetic field has three independent
components $F= F_{xy} dx dy + F_{yz} dy dz + F_{zx} dz dx$. Any
proper rotation has at least one unit eigenvalue and two complex
conjugate ones of unit norm. By elementary physics the
unmagnetized direction points along the magnetic field $\vec{F}=
(F_{yz}, F_{zx}, F_{xy})$. Setting $f=|\vec{F}|$, the modes of the
coordinates orthogonal to $\vec{F}$ of a singly charged open
string, starting on a magnetic brane and ending on an unmagnetized
one, are shifted by \be \epsilon = {\beta\over \pi } \quad {\rm
with} \quad \beta = \arctan(f) \quad .\ee The situation is
slightly more involved for a string starting on a magnetized brane
labeled by $a$ and ending on another one labeled by $b$. As
anticipated in the introduction and suggested by boundary state
considerations \cite{divelic}, one has to diagonalize the rotation
matrix $R_{ab} = R_a^{q_a}R_b^{q_b}$, where $R_a$, ($R_b$) is a
rotation around $\vec{F}_a$ ($\vec{F}_b$) by an angle $2\beta_a$
($2\beta_b$). The eigenvector corresponding to the unit eigenvalue
identifies the unmagnetized direction $\vec{u}_{ab}$, while the
two orthogonal directions generically suffer a non-trivial action
of $R_{ab}$. For practical reasons, instead of working in the
3-dimensional vector representation of $SO(3)\approx SU(2)$, it
turns out to be much more rewarding to work in the spin 1/2
representation and exploit the remarkable properties of Pauli
matrices. A magnetic rotation can be written as \be R^q_{(1/2)} =
\cos(\beta) - i q \hat{F}\cdot \sigma \sin(\beta)  \ee where
$\hat{F}= \vec{F}/|\vec{F}|$ is the unit vector in the direction
of the magnetic field. Let us stress once more that the argument
of the trigonometric functions is correctly $\beta = 2\beta/2$.
Denoting for short $\cos(\beta_{a/b})$, respectively
$\sin(\beta_{a/b})$, by $c_{a/b}$, respectively $s_{a/b}$, and
combining the two rotations yields \be R_{ab} = (c_a c_b - q_a q_b
s_a s_b \hat{F}_a \cdot \hat{F}_b) - i (q_a s_a c_b \hat{F}_a +
q_b c_a s_b \hat{F}_b + q_a q_b s_a s_b \hat{F}_a \times
\hat{F}_b)\cdot \sigma \ee that exposes the `unmagnetized'
direction \be \vec{u}_{ab} = q_a s_a c_b \hat{F}_a + q_b c_a s_b
\hat{F}_b + q_a q_b s_a s_b \hat{F}_a \times \hat{F}_b \ee and
determines the magnetic shifts $ \epsilon_{ab}=\beta_{ab}/\pi$ via
\be \cos(\beta_{ab}) = c_a c_b - q_a q_b s_a s_b \hat{F}_a \cdot
\hat{F}_b \quad .\ee

Though obvious, let us stress once more that $\beta_{ab}\neq
\beta_{a}\pm \beta_{b}$ in general, contrarily to the
two-dimensional as bad as any factorizable case. Another amusing
feature is the orientation of the `unmagnetized' direction
$\vec{u}_{ab}$  with respect to the generators of the fundamental
cell. Due to Dirac quantization condition on ${F}_{a}$, in each
sector of the spectrum one can redefine the fundamental cell and
choose $\vec{u}_{ab}$ to be one of the generators thereof. The
generalized KK momenta carried by neutral strings turn out then to
be rather involved combinations of momenta and windings. These are
not so surprising features of the non-abelian structure of the
rotation group.

Unfortunately, there is no non-trivial way to preserve
supersymmetry and, as in the two dimensional case, one can neither
avoid tachyons when non-trivial magnetic fields are turned on, nor
stabilize the vacuum perturbatively.

{} {}{}
{} {}
{{}}
{{{}}}
\section{Four-dimensional magnetized torus}
\label{tquat}

Magnetized four-dimensional tori display some remarkable features.
This is the lowest dimensional case where non-trivial
supersymmetric configurations first appear. Moreover thanks to the
non-simplicity of the rotation group, $SO(4)=SU(2)_L \times
SU(2)_R$, one can compactly analyze the problem for arbitrary
magnetic fields. To this end it is very convenient to choose a
basis for the six constant two-forms comprising three self-dual
$\omega^a$ and as many anti self-dual $\bar\omega^a$ two-forms and
decompose $F$ according to \be F = {1\over 2} F_{ij} dx^i dx^j =
F^+ + F^- = f_r \omega^r + \bar{f}_r \bar\omega^r \quad .\ee

This can be rather explicitly achieved, for instance, by using 't
Hooft symbols $\eta^a_{mn}$ and $\bar\eta^r_{ij}$, with $r=1,2,3$
and $i,j=1,...4$, and setting $\omega^r = \eta^r_{ij} dx^i dx^j/2$ and
$\bar\omega^r =\bar\eta^r_{ij}dx^i dx^j/2$. 't Hooft symbols are
real and antisymmetric and satisfy \bea
\eta^r_{i}{}^{k}\eta^s_{k}{}^{j} &&= -
{\delta}^{rs}{\delta}_{i}{}^{j}
+ \epsilon^{rs}{}_{t}\eta^t_{i}{}^{j} \nonumber\\
\bar\eta^r_{i}{}^{k}\bar\eta^s_{k}{}^{j} &&= -
{\delta}^{rs}{\delta}_{i}{}^{j} +
\epsilon^{rs}{}_{t}\bar\eta^t_{i}{}^{j}
\nonumber\\
\eta^r_{i}{}^{k}\bar\eta^s_{k}{}^{j} && =
\bar\eta^s_{i}{}^{k}\eta^r_{k}{}^{j} \eea In terms of Pauli
matrices it is easy to check that \bea &&\eta^1 = i\sigma_2
\otimes \sigma_1 \quad \eta^2 = i\sigma_2 \otimes \sigma_3 \quad
\eta^3 = 1 \otimes i\sigma_2 \nonumber \\
&&\bar\eta^1 = \sigma_1 \otimes  i\sigma_2  \quad
\bar\eta^2 = \sigma_3 \otimes  i\sigma_2 \quad
\bar\eta^3 = i\sigma_2 \otimes 1
\eea do the job.

For singly charged strings, the magnetic rotation matrices, whose
eigenvalues determine the magnetic shifts, assume the form \be R =
(1 - H) (1+ H)^{-1}= (1 - H^+ - H^-) (1+ H^+ + H^-)^{-1} = R_L R_R
= R_R R_L\ee where $H^i{}_{j} = G^{ik}F_{kj}$. After
straightforward algebra, that heavily relies on the use of $[H^+,
H^-] =0$, $(H^+)^2 = - |h|^2\identmat$ and $(H^-)^2 = -
|\bar{h}|^2\identmat$, with $|v|^2\equiv \sum_r v_r^2$, one finds
\bea && R_L = {{ (1-|h|^2 + |\bar{h}|^2) - 2 H^+}\over\sqrt{1 +
2(|h|^2 + |\bar{h}|^2) + (|h|^2 - |\bar{h}|^2)^2 }} \\ &&R_R = {{
(1+|h|^2 - |\bar{h}|^2) - 2 H^-}\over\sqrt{1 + 2(|h|^2 +
|\bar{h}|^2) + (|h|^2 - |\bar{h}|^2)^2 }} \quad .\eea The four bosonic
string coordinates transform in the $(1/2,1/2)$ representation of
$SO(4)=SU(2)_L \times SU(2)_R$, and thus the magnetic shifts of the
modes of singly charged spacetime bosons, with one end on an
unmagnetized stack, are determined by the two pairs \be
\beta_{Q=1}  = \pm (\beta_L \pm \beta_R) \ee with \be \tan (2
\beta_L) = {{ 2 |h|} \over {1 - |h|^2 + |\bar{h}|^2} } \quad ,
\quad \tan (2 \beta_R) = {{ 2 |\bar{h}|} \over {1 + |h|^2 -
|\bar{h}|^2} } \label{singly} \quad . \ee

Spacetime spinors transforming in the $(1/2,0)$, respectively
$(0,1/2)$, representation have their masses shifted by
$\pm\beta_L/\pi$, respectively $\pm \beta_R/\pi$. Thus the only
way to preserve some susy in this (singly charged) sector is
having either $\beta_L = 0$ or $\beta_R = 0$, which is tantamount
to turning on only (anti) self-dual magnetic fields, \ie abelian
(anti)instantons on ${\bf T}^4$. 

In the self-dual case, choosing as reference complex structure,
\eg $J_{ij} =\bar{\eta}^3_{ij}$, \ie setting $z^1 = x^1 + i x^2$
and $z^2 = x^3 + i x^4$, shows that the magnetic two-form is a
type $(1,1)$ form, \ie $F_{(2,0)}=0$, with $F_{(1,1)}\wedge J =0$. In
mathematical terms, one is dealing with a hermitean connection on
a holomorphic stable bundle. The choice of complex coordinates,
though particularly suitable to exposing susy and reducing the
problem to a two-dimensional one, hides the beautiful
hyperk\"ahler structure coded in the ADHM construction of
instantons on ${\bf T}^4$ or on its susy preserving orbifolds
${\bf T}^4/\Gamma \approx {\bf K}3$.

For doubly charged strings, the situation is analogous and one gets
\be
\beta_{Q=2} = \pm 2 (\beta_L \pm \beta_R) \label{doubly}
\ee
for spacetime bosons and $\beta_{Q=2}=\pm 2\beta_L$ or
$\beta_{Q=2}=\pm 2\beta_R$ for spacetime fermions.

The situation is slightly more involved for open strings ending on
branes with different magnetic fields, \ie carrying charge
$q_a=\pm 1$ with respect to the $U(1)_a$ subgroup associated to
the first stack of $N_a$ branes and charge $q_b = \pm 1$ with
respect to the $U(1)_b$ subgroup associated to the second stack of
$N_b$ branes. Yet exploiting the close analogy between 't Hooft
symbols, or rather the $4\times 4$ matrices $\Sigma^a = i \eta^a$
and $\bar\Sigma^a = i \bar\eta^a$, and Pauli matrices $\sigma^a$,
it takes a few minutes thinking to convince oneself that the
relevant matrices \be R_{abL} = R_{aL}^{q_a}R_{bL}^{q_b} \quad ,
\quad R_{abR} = R_{aR}^{q_a}R_{bR}^{q_b} \ee that determine
$R_{ab} = R_{abL} R_{abR} = R_{abR} R_{abL}$ can be conveniently
expressed in the form \be R_L^q = \cos(2\beta_L) - i q
\sin(2\beta_L) \hat{h}\cdot \Sigma \quad , \quad R_L^q =
\cos(2\beta_R) - i q \sin(2\beta_R) \hat{\bar{h}}\cdot \bar\Sigma
\ee where $\tan(2\beta_{L/R})$ have been defined above and a hat
over a three-component vector means the unit vector in its
direction. Denoting for short $\cos(2\beta_{a,b})$, respectively
$\sin(2\beta_{a,b})$, by $c_{a,b}$, respectively $s_{a,b}$, and
combining the two rotations yields \be R_{abL} = (c_a c_b - q_a
q_b s_a s_b \hat{h}_a \cdot \hat{h}_b) - i (q_a s_a c_b \hat{h}_a
+ q_b c_a s_b \hat{h}_b + q_a q_b s_a s_b \hat{h}_a \times
\hat{h}_b)\cdot \Sigma \ee and \be R_{abR} = (\bar{c}_a \bar{c}_b
- q_a q_b \bar{s}_a \bar{s}_b \hat{\bar{h}}_a \cdot
\hat{\bar{h}}_b) - i (q_a \bar{s}_a \bar{c}_b \hat{\bar{h}}_a +
q_b \bar{c}_a \bar{s}_b \hat{\bar{h}}_b + q_a q_b \bar{s}_a
\bar{s}_b \hat{\bar{h}}_a \times \hat{\bar{h}}_b)\cdot \bar\Sigma
\ee The magnetic shifts of the modes of the four bosonic
coordinates (two complex pairs) \be \beta_{ab} = \pm \beta_{abL}
\pm \beta_{abR} \ee are determined by \be \cos(2\beta_{abL}) = c_a
c_b - q_a q_b s_a s_b \hat{h}_a \cdot \hat{h}_b \quad , \quad
\cos(2\beta_{abR}) = \bar{c}_a \bar{c}_b - q_a q_b \bar{s}_a
\bar{s}_b \hat{\bar{h}}_a \cdot \hat{\bar{h}}_b \quad .
\label{doublyab}\ee
Equivalent but less explicit formulae have been found in \cite{lars}.

Generically, there are no unmagnetized directions. Indeed
$\beta_{ab}\neq 0$ unless ${h}_a = \pm \bar{h}_a$, in which case
one is effectively restricting the fluxes to a three-dimensional
torus and considering the diagonal $SO(3)$ subgroup. Strings
connecting branes of different kinds may however appear with
non-trivial multiplicities associated to the presence of fermionic
zero-modes counted by the index of the Dirac operator coupled with
charge $q_a$ to the magnetic field $F_a$ on the first stack of
$N_a$ branes and with charge $q_b$ to the magnetic field $F_b$ on
the second stack of $N_b$ branes. The index vanishes in odd
dimension. In four dimensions it is given by the second Chern
number $c_2(q_a {\cal F}_a + q_b {\cal F}_b)$ of the abelian gauge
bundle \bea I_{ab} &=& - {1 \over 2 (2\pi)^2} \int_{{\Sigma}^4}
Tr_{(N_a,N_b)}[(q_a {\cal F}_a + q_b {\cal F}_b) \wedge
(q_a {\cal F}_a + q_b {\cal F}_b)] \nonumber \\
&=& {W_a W_b \over (2\pi)^2} [- |q_a h_a + q_b h_b|^2 + |q_a
\bar{h}_a + q_b \bar{h}_b|^2] \quad ,\label{index} \eea where
${\cal F}_{a/b}$ are the world-volume magnetic fields and $W_{a/b}
= \det(W_{a/b})$ take care of the oriented brane wrappings so that
$N_{a/b} = \hat{N}_{a/b} |W_{a/b}| $. For ``parallel" magnetic
fields in factorized rectangular two-tori ${\bf T}^4 = {\bf
T}^2_{(1)}\times {\bf T}^2_{(2)}$, $W_{a/b}=\prod_{i=1}^2
n^{(i)}_{a/b}$ and (\ref{index}) reduces to the simpler expression
\be I_{ab} = {W_a W_b \over (2\pi)^2} \prod_{i=1}^2(q_a h^{(i)}_a
+ q_b h^{(i)}_b) = \prod_{i=1}^2 (q_a m^{(i)}_a n^{(i)}_b + q_b
m^{(i)}_b n^{(i)}_a) \ee upon imposing Dirac quantization
condition on each subtorus and for each stack of branes. In
summary, there are ${I}_{ab}$ strings in the representation
$(\hat{N_a}, \hat{N_b}^*)$ or $(\hat{N_a}, \hat{N_b}^*)$, present
only in the unoriented case, of the Chan-Paton group commuting
with the magnetic $U(1)$'s. The analysis of the doubly charged
strings, present only in the unoriented case, is similar. In order
to correctly determine the multiplicities one has to properly take
into account the action of $\Omega$ that (anti) symmetrizes the
Chan-Paton indices.

Although we have all the necessary ingredients to write down the
one-loop partition function for a magnetized four-torus, we
refrain from doing so in order to avoid cumbersome formulae that
can be conveniently  subsumed by the following observations. As in
the two-dimensional case, one has to distinguish several sectors.

Neutral strings with $q_a=q_b=0$, starting and ending on
unmagnetized branes, suffer neither magnetic shifts of their oscillator modes
nor rescalings of their KK momenta.

Dipole strings with $q_a=-q_b=\pm 1$, starting and ending
on the same stack of magnetized branes, suffer no magnetic shifts but
carry rescaled KK momenta.

Singly charged strings $q_a=\pm 1$ and $q_b=0$ or {\it vice
versa}, starting on a stack of magnetized branes and
ending on a stack of unmagnetized branes, suffer magnetic shifts given by
(\ref{singly}) and, generically, carry no KK momenta but rather multiplicities
determined by the index theorem (\ref{index}) to be $c_2({\cal F}_a)$.

Doubly charged strings with $q_a=q_{\tilde{a}}=\pm 1$, starting  on
magnetized branes and ending on their images under $\Omega$, suffer
double magnetic shifts and,
generically, carry no KK momenta but rather multiplicities determined by
the index theorem to be $c_2(\pm 2{\cal F}_a)= \pm 4 c_2({\cal F}_a)$, up to
(anti)symmetrization imposed by $\Omega$.

Dy-charged strings with $q_a=\pm 1$ and $q_b=\pm 1$,
starting on a stack of magnetized branes and ending on a different
stack of magnetized branes or their images under $\Omega$, suffer
magnetic shifts determined by the trigonometric formula (\ref{doublyab}),
generically carry no KK momenta but
rather discrete multiplicities determined by the index theorem, \ie
$c_2(q_a {\cal F}_a + q_b {\cal F}_b)$ up to identifications resulting
from $\Omega$.

So far we have tacitly assumed a square torus and vanishing NS-NS
antisymmetric tensor background. For an arbitrary torus one needs
to replace the coordinate differentials $dx^i$ with the tetrad
$E^\alpha = E^\alpha_i dx^i$ so that $G_{ij} =
\delta_{\alpha\beta} E^\alpha_i  E^\beta_j $. This may not be
enough in the case of non-trivial wrapping whereby
$W_{(a)\alpha}^i\neq \delta^i_\alpha$ for some $a$. In the absence
of magnetic field the integers $W_{a} = \det(W_\alpha{}^i_{(a)})$
simply rescale the tension of the stack of branes $\tau_{a} = \tau
N_{a}|W_{a}|$ and their R-R charge density $\rho_{a} = \tau
N_{a}W_{a}$, so that the stacks preserve the same 16 supercharges
only when $W_{a}W_{b}>0$ for all $a$ and $b$. Moreover in type I
theories this fraction coincides with the one preserved by
standard $\Omega$-planes for a given choice of sign, let us say
$W_{a}>0$. For vanishing NS-NS antisymmetric tensor background,
R-R tadpole cancellation requires $2 \sum_a N_{a}W_{a} = 32$,
after including image branes. The situation becomes more
interesting when some magnetic fields are turned on. The subtle
point is that the magnetic field that satisfies Dirac quantization
condition is the worldvolume magnetic field ${\cal
F}^a_{\alpha\beta}$ rather than $F_{ij}$, that need not have
integer `periods', \ie first Chern numbers $c_{ij}$. Rather
$c_{\alpha\beta} = W_\alpha{}^i W_\beta{}^j c_{ij}$ are integrally
quantized, except possibly for a half-integer shift related to the
presence of (the pull-back of) a quantized NS-NS antisymmetric
tensor background ${\cal B}_{\alpha\beta} = W_\alpha{}^i
W_\beta{}^j B_{ij}$. As observed by \cite{raba}, this may not be
sufficient. In general, one has to require integrality of all
Chern numbers, which are T-dual to all possible (sub)brane
charges. In practise, one has to solve the inverse problem, \ie
given rational numbers $c^1_{ij}$, $c^2_{ij,kl}$, etc, one should
find an integer matrix $W_\alpha{}^i$ such that
$c^1_{\alpha\beta}$, $c^2_{\alpha\beta,\gamma\delta}$, etc are
(half) integer compatibly with R-R tadpole conditions. This may
fail with purely abelian fields. In some cases \cite{raba}, adding
a non-abelian part might help imposing integrality but would
require abandoning the exact worldsheet analysis of the present
paper.

We are now ready to discuss the potential for the closed string
moduli  and address the issue of their stabilization. The
potential is generated by the Born-Infeld lagrangian \cite{leigh}
\be \sqrt{\det({\cal G}_{\alpha\beta}+{\cal F}_{\alpha\beta})} =
\sqrt{\det[W_\alpha{}^i W_\beta{}^j (G_{ij} + F_{ij})]} =
|\det(W_\alpha^i)| \sqrt{\det(G_{ij} + F_{ij})}\ee and the WZ
coupling to the R-R fields \bea &&\int_{\Sigma^d} \sum_k {\cal
C}_{k+1} Tr(e^{\cal F}) = \sum_a W_a \left[ C_{10-d} \int_{{\bf
T}^d} {\omega}_d N_a +  \sum_f C^f_{8-d+2}\int_{{\bf T}^d}
{\chi}^r_{d-2} \wedge{F_a} \right.\nonumber \\ &&\left.+ \sum_p
C^q_{6-d+2} \int_{{\bf T}^d} {\psi}^q_{} \wedge{F_a}\wedge{F_a} +
\sum_u C^u_{4-d+2} \int_{{\bf T}^d} {\varphi}^u_{} \wedge{F_a}
\wedge{F_a}\wedge{F_a} \right]\eea Terms with an odd number of
magnetic fields disappear upon $\Omega$-projection. Notice that
once $\det(W_\alpha^i)$ is factored out it is the non-integrally
(but rationally) quantized $F_a$ that appears. The Born-Infeld
term \cite{leigh} yields \be \sqrt{\det({\cal G}_{ij}+ {\cal
F}_{ij})} = \sqrt{\det({\cal G}_{ij})} \sqrt{\det(\identmat +
{\cal H})} \ee where ${\cal H}^i{}_j = {\cal G}^{ik} {\cal
F}_{kj}$. In turn \be \det(\identmat + {\cal{H}}) =
\exp\left[tr(\log(\identmat + {\cal{H}})\right] =
\exp[\sum_{k=0}^{[d/2]}{1\over 2k} tr({\cal{H}}^{2k})] \ee where
$tr(..)$ denote a trace over target space vector indices, not to
be confused with $Tr(..)$ denoting a trace over Chan-Paton
indices. Indeed it is easy to show that $tr({\cal{H}}^{2k+1})=0$.
Similarly the restriction to $k<[d/2]$ can be reached via group
theory considerations as well as inspection of the initial
determinant that can expose terms with at most ${\cal F}^d$.

More explicitly, for $d=2,3$
\be
{\det({\cal G} + {\cal F}) \over \det({\cal G})} =  [1 - {1\over 2}
tr({\cal{H}}^{2})] = (1 + h^2)
\label{h23}
\ee
where $h$ is the skew eigenvalue of ${\cal{H}}$.

Similarly, for $d=4,5$, one has
\be
{\det({\cal G} + {\cal F}) \over \det({\cal G})}= [1 - {1\over 2}
tr({\cal{H}}^{2}) - {1\over 4} tr({\cal{H}}^{4}) + {1\over 8}
(tr({\cal{H}}^{2}))^2] = (1 + h_1^2)(1 + h_2^2)
\label{h45}
\ee
where $h_i$ with $i=1,2$ are the two skew eigenvalues of ${\cal{H}}$.

Finally, for $d=6,7$, one finds
\bea
&&{\det({\cal G} + {\cal F}) \over \det({\cal G})}=
[1 - {1\over 2} tr({\cal{H}}^{2}) - {1\over 4}
tr({\cal{H}}^{4}) + {1\over 8} (tr({\cal{H}}^{2}))^2 -
\nonumber \\
&& \quad\quad- {1\over 6} tr({\cal{H}}^{6}) + {1\over
8}tr({\cal{H}}^{2}) tr({\cal{H}}^{4}) - {1\over 48}
(tr({\cal{H}}^{2}))^3] = \prod_{i=1}^3 (1 + h_i^2)
\label{h67}
\eea
where $h_i$ with $i=1,2,3$ are the skew eigenvalues of ${\cal H}$.

Let us now specialize to $d=4$, where one can exploit the
hyperk\"ahler structure of ${\bf T}^4$ and decompose ${\cal{H}}$
into its self-dual and anti-self-dual parts ${\cal{H}}=
{\cal{H}}_+ + {\cal{H}}_-$. Then one can rewrite (\ref{h45}) as
\bea &&1 - {1\over 2} tr({\cal{H}}^{2}) - {1\over 4}
tr({\cal{H}}^{4}) + {1\over 8} (tr({\cal{H}}^{2}))^2 = \\&&1 -
{1\over 2} [tr({\cal H}_+^{2}) + tr({\cal H}_-^{2})] + {1\over 16}
[tr({\cal H}_+^{2}) - tr({\cal H}_-^{2})]^2 \quad .\nonumber\eea
Combining with the volume factor, the Born-Infeld lagrangian of
each stack of branes can be elegantly written in the form \bea &&
\sqrt{\det({\cal G})} \left[ (1 \mp {1\over 2} [tr({\cal H}_+^{2})
- tr({\cal H}_-^{2})])^2 +
tr({\cal H}_\mp^{2})\right]^{1/2} = \nonumber \\
&& |W| \left[\left({1\over 2} \int J\wedge J - {1\over 2} \int
F\wedge F\right)^2 + \left(\int J\wedge F\right)^2 + \vert\int
\Omega\wedge F\vert^2 \right]^{1/2}\eea where $\Omega = J^1 + i
J^2= J^+$ is the complex (2,0)-form and $J= J^3$ is the K\"ahler
form in the reference complex structure. Our normalizations are
such that $\Omega\wedge \bar\Omega = 2 J\wedge J$. At the linear
level the susy preserving BPS conditions require that either
${\cal H}_+=0$ or ${\cal H}_-=0$ for all magnetic $U(1)$'s. In the
latter case, ${\cal H}_-=0$, the last three terms drop out and one ends
up with $$ \sqrt{\det({\cal G})} \vert 1 - {1\over 2} tr({\cal
H}_+^2)\vert = \sqrt{\det({\cal G})} \left( 1 - {1\over 2}
tr({\cal H}_+^2)\right) \quad . $$ Indeed, both terms are positive
definite since ${\cal H}_+$ being real and antisymmetric is
anti-hermitean. The same is true in the former case,  ${\cal
H}_+=0$, after reversing the orientation of ${\bf T}^4$.

Contrary to the 2- and 3-dimensional cases, the 1/2-loop potential
for the NS-NS moduli induced by the Born-Infeld term combined with
the contribution of $\Omega$-planes depends {\it a priori} on all
the moduli. Rescaling the metric according to $\widehat{G} =
{G}/\det(G)^{1/4}$, it can be cast into the form \be
V(\phi,\omega, \hat{G}; {\cal F}_a) = \tau e^{-\phi} \left( 2
\sum_a N_a \sqrt{|W_a|^2\omega^2 + |W_a|\omega c^2_1({\cal F}_a) +
c_2^2({\cal F}_a)} \mp 32 |\omega| \right) \label{pot4} \ee where
$\omega = \sqrt{\det(G)}$ parametrizes the overall volume of ${\bf
T}^4$. Depending on whether $c^2_1({\cal F}_a)$ is larger or
smaller than $2|c_2({\cal F}_a)|$ and on the sign of the
$\Omega$-plane tension one gets different behaviours for $\omega$.
For supersymmetric configurations $c^2_1({\cal F}_a)= \pm 2
c_2({\cal F}_a)$ (the same sign for all $a$) and one simply gets
\be V(\phi,\omega, \hat{G}; F_a) = \tau e^{-\phi} \left( 2 \sum_a
N_a (|W_a| |\omega| + |c_2({\cal F}_a)|) \mp 32 |\omega| \right)
\label{pot4} \ee this should be supplemented with the R-R tadpole
conditions $2 \sum_a N_a W_a = \pm 32$ (for $C_{10}$) and $2
\sum_a N_a W_a c_2(F_a) = 0$ (for $C_6$) that cannot be satisfied
for non-trivial magnetic fields unless one introduces lower
dimensional $\Omega$-planes.

For non supersymmetric configurations, R-R tadpoles can be
satisfied and in principle all the closed string moduli, including
$\omega$ yet excluding $\phi$, can be stabilized in the presence
of enough stacks of  branes with oblique magnetic fluxes. One
cannot prevent however the presence of open string tachyons that
destabilize the vacuum anyway.

We would like to conclude this section with an amusing observation
on non-linear self-duality \cite{mmms,dt}. As discussed above,
barring the overall volume factor, the Born-Infeld action in $d=4$
can be compactly written as \be \sqrt{\det(1 + {\cal H})} =
\sqrt{(1+h_1^2)(1+h_2^2)} = \sqrt{(1\pm h_1 h_2)^2 + (h_1\mp
h_2)^2} \label{BI4} \ee where $h_i$ with $i=1,2$ are the two skew
eigenvalues of $H$. ``Standard'' (anti) self-dual and thus
supersymmetric configurations correspond to $h_1\mp h_2 = 0$ and
yield \be \sqrt{\det(1 + {\cal H})} = |1\pm h_1 h_2| = 1 + |h|^2
\label{BI42} \ee that, combined with the R-R coupling, disclose
the presence of BPS D5- or $\bar{\rm D5}$-branes inside the
magnetic D9-branes \cite{braninbran}. The resulting 1/4 BPS bound
state is at threshold since the positive tensions simply add. In
order to `dispose of' the square root in (\ref{BI4}) and find
other bound states at threshold one can however consider the case
$1\pm h_1 h_2 =0$, \ie $h_1 =\pm 1/h_2$ that does not lead to
standard BPS branes but corresponds to a non (manifestly) BPS
state which is T-dual to a BPS configuration of D5-branes bound to
pairs of D9- and $\bar{\rm D9}$-branes. It is natural to ask to
what extent the initial configuration satisfying the non-linear
(anti) self-duality condition is non-BPS. It is not unconceivable
that BPS-ness can be regained, even before perfoming any
T-duality, with respect to local supersymmetries associated to
gravitini in the closed string spectrum that become massless at
the non-linear (anti) self-dual point.

{} {}{}
{} {}
{{}}
{{{}}}
\section{Five-dimensional magnetized torus}
\label{tcinq}

A constant magnetic field $F$ in five dimensions has  10
independent components. Four of them may be taken to be associated
to the independent components of the unit vector $\vec{u}$ along
the "unmagnetized" direction associated to the unit eigenvalue of
the rotation matrix $R(F)$. The other six
determine the magnetic field in the four transverse generically
``magnetized" directions. By means of this splitting one can easily
determine the magnetic shifts for singly and doubly charged
strings and the rescaling of the momenta for dipole strings. The
study of dy-charged strings, although straightforward in
principle, is quite laborious in practice since one has to
diagonalize a generic $5\times 5$ orthogonal matrix, something
that leads to very cumbersome formulae. Yet, the strategy is first
to identify the ``unmagnetized" direction by solving $R_{ab}
\vec{u}_{ab} = \vec{u}_{ab}$, then decompose the reduced $4\times
4$ orthogonal matrix according to $R^{\perp}_{ab}= R^{\perp
L}_{ab}R^{\perp R}_{ab}$ and finally exploit the properties of 't
Hooft symbols in order to determine the eigenvalues and the resulting magnetic shifts.

The condition for supersymmetry in $D=5$ are as stringent  as in
$D=6$. In concrete terms the magnetic rotation matrices should all
belong to the same $SU(2)=Sp(2)$ subgroup of $SO(5)=Sp(4)$. In
particular this means that they should all have the same
unmagnetized direction $\vec{u}$, which may or may not coincide
with one of the generators of the fundamental cell. The holonomy
of the $SO(4)$ stability group should further reduce to $SU(2)$.

Since nothing essentially new can be learned from this case  we do
not go into any further detail.

{} {}{}
{} {}
{{}}
{{{}}}
\section{Six-dimensional magnetized torus}
\label{tsei}

Six dimensional magnetized tori represent a genuinely new  and
phenomenologically interesting case. In general, a magnetic field
has 15 independent components, corresponding to the constant and
thus closed two-forms. For obvious (supersymmetry) purposes it is
convenient to choose complex coordinates and set \be F^a =
F^a_{(2,0)} + F^a_{(1,1)} + F^a_{(0,2)} \ee where
$F^a_{(2,0)}$,
$F^a_{(1,1)}$ and $F^a_{(0,2)}$ have 6, 9 and 6 independent
components respectively. Notice that $F^a_{(0,2)} =
F^{a*}_{(2,0)}$, $ F^a_{(1,1)} = F^{a*}_{(1,1)}$ for a real
gauge field, \ie a hermitean connection.

In order to preserve some supersymmetry \cite{mmms,am,dt}, the
complex structure of ${\bf T}^6$ must be chosen so that for all
$a$ \be F^a_{(0,2)} = 0 \ee and the K\"ahler form $J= i G_{i
\bar{j}} dz^i \wedge dz^{\bar{j}}$ must satisfy \be {1\over 2!}
J\wedge J \wedge F^a_{(1,1)} = {1\over 3!} F^a_{(1,1)}\wedge
F^a_{(1,1)} \wedge F^a_{(1,1)} \quad , \label{stabil} \ee \ie the
magnetic bundle must be holomorphic and stable wrt to $J$. In turn
this implies a reduction of the holonomy group generated by the
magnetic rotation matrices $R_a$ from $SO(6)= SU(4)$ to $SU(3)$
for ${\cal N}=1$ supersymmetry or $SU(2)$ for ${\cal N}=2$
supersymmetry. ${\cal N}=4$ supersymmetry is enjoyed by the
neutral open strings, for which $R_0 =\identmat$.

In the supersymmetric case, the $6\times 6$  rotation matrices are
direct sums of $3\times 3$ blocks \be R =\left(
\begin{array}{cc} U_{\bf 3}  & 0 \\ 0 & U_{\bf 3^*}\ea\right)
\ee since ${\bf 6} \rightarrow {\bf 3} + {\bf 3^*}$ under
$SU(4)\rightarrow U(3)$. More
explicitly \be U_{\bf 3}^i{}_j = (\delta^i{}_k - i H^i{}_k)
(\delta^k{}_j + i H^k{}_j)^{-1} \ee where $H^i{}_j =
G^{i\bar{k}}F_{\bar{k} j}$ and similarly for $U_{\bf 3^*}$.

The stability condition is not simply $H^i{}_i=0$, as in $d=4$
where it arose from $J\wedge F =0$, but rather the condition for
$SU(3)$ holonomy: $\det(U_{\bf 3})=1$.  In other words $\det(1+iH)
= \det(1-iH)$ that implies$\sum_k (-)^k tr(H^{2k+1})/(2k+1)=0$.
Since in general $tr_{\bf 3}(H)\neq0$, one can expand $H_{\bf 3}$
in  the basis of generators of $U(3)$ \be H = h_0 \identmat + h_r
\lambda^r \ee where we choose the eight $3\times 3$ Gell-Mann
matrices $\lambda^r$ so as to be hermitean and satisfy \be
tr(\lambda^r \lambda^s) = 3 \delta^{rs} \qquad \lambda^r \lambda^s
= \delta^{rs} \identmat + (i f^{rst} + d^{rst})\lambda_t \quad ,
\label{gellmann}\ee where the $SU(3)$ structure constants
$f^{rst}$ are totally  antisymmetric and the anomaly coefficients
$d^{rst}$ are totally symmetric.

In order to determine the eigenvalues $\rho_k = \exp(2 i \beta_k)$
of $U$ in terms of the 9 components $h_0$ and $h_r$ of $H$ it is
convenient to diagonalize $\hat{H}= h_r \lambda^r = h \cdot
\lambda$ so that \be \rho_k = {1 - i h_0 -i \mu_k \over 1 + i h_0
+i \mu_k} \ee Using $tr (\hat{H}) = 0$, $tr (\hat{H}^2) = 3
\delta^{rs}h_r h_s \equiv 3|h|^2$ and $tr( \hat{H}^3) = 3
d^{rst}h_r h_s h_t \equiv 3 dhhh$, the secular equation for
$\hat{H}$ boils down to a cubic in canonical form \be \mu^3 -
{3\over 2}|h|^2 \mu - dhhh =0 \quad ,\ee whose solutions are given
by Cardano - Tartaglia formula \bea \mu_k &=& {e^{2k\pi i/3}\over
2^{1/3}} \left[dhhh + i \sqrt{{1\over
2}|h|^6 - (dhhh)^2} \right]^{1/3} \nonumber \\
&+& {e^{-2k\pi i/3}\over 2^{1/3}} \left[dhhh  - i \sqrt{ {1\over
2}|h|^6 - (dhhh)^2} \right]^{1/3} \quad , \label{eigenh}\eea where
$|h|^6 \ge 2 (dhhh)^2$ has been used.
 Supersymmetry
translates into the condition \be h_0^3 + h_0^2 \sum_k \mu_k + h_0
( \sum_{i<j}\mu_i\mu_j - 3) + \prod_k \mu_k - \sum_k \mu_k = 0
\quad . \ee

We now consider the more laborious case of strings ending  on
different stacks of branes, whose magnetic shifts are determined
by the eigenvalues of the $SU(3)$ matrix  $U_{ab} = U_a^{q_a}
U_b^{q_b}$. Setting for definiteness $q_a = 1$ and $q_b=-1$ and
factoring out an overall $U(1)$ phase gives \bea U_{ab} &=& {(1 -
i h^0_a)(1 + i h^0_b) \over (1 + i h^0_a)(1 - i h^0_b) } (1 - i
\hat{h}_a\cdot \lambda)(1 + i \hat{h}^*_a \cdot\lambda)^{-1} (1 -
i \hat{h}^*_b\cdot \lambda)(1 + i
\hat{h}_b\cdot \lambda)^{-1} \nonumber \\
&=& {(1 - i h^0_a)(1 + i h^0_b) \over (1 + i h^0_a)(1 - i h^0_b) }
\ \hat{U}_{ab} \eea where $\hat{h}_{a/b}= {h}_{a/b}/(1 - i
h^0_{a/b})$ and $*$ denotes complex conjugation. Perusing
(\ref{gellmann}) yields \be \hat{U}_{ab} = M_a^{-1} {1 - \hat{h}_a
\cdot \hat{h}_b\over 1 - \hat{h}^*_a \cdot \hat{h}^*_b} (1 - i
\hat{h}_{ab} \cdot \lambda) (1 + i\hat{h}^*_{ab} \cdot
\lambda)^{-1} M_a \ee where $M_a =  1 + i \hat{h}^*_a
\cdot\lambda$ and \be \hat{h}_{ab} = {\hat{h}_{a} + \hat{h}_{b} +
f \hat{h}_{a} \hat{h}_{b} - i d \hat{h}_{a} \hat{h}_{b} \over 1 -
\hat{h}_a \cdot \hat{h}_b } \ee so that \be \hat{\mu}_{ab,k} = {1
- \hat{h}_a \cdot \hat{h}_b\over 1 - \hat{h}^*_a \cdot
\hat{h}^*_b} \hat{\nu}_{ab,k} \ee where \bea \hat{\nu}_{ab,k} &=&
{e^{2k\pi i/3}\over 2^{1/3}}
\left[d\hat{h}_{ab}\hat{h}_{ab}\hat{h}_{ab}  + i \sqrt{{1\over
2}|\hat{h}_{ab}|^6 - (d\hat{h}_{ab}\hat{h}_{ab}\hat{h}_{ab})^2 }
\right]^{1/3} \nonumber \\ &+& {e^{-2k\pi i/3}\over 2^{1/3}}
\left[d\hat{h}_{ab}\hat{h}_{ab}\hat{h}_{ab}  - i \sqrt{ {1\over
2}|\hat{h}_{ab}|^6 - (d\hat{h}_{ab}\hat{h}_{ab}\hat{h}_{ab})^2  }
\right]^{1/3} \eea and eventually one can compute ${\rho}_{ab,k} =
\exp(2 i \beta_{ab,k})$ and get the magnetic shifts
$\epsilon_{ab,k}= \beta_{ab,k}/\pi$, where $k=1,2,3$ labels the
three orthogonal directions along the eigenevectors of $U_{ab}$.
Clearly the eigenvalues of $R$ come in  complex conjugate pairs
$(\rho_k, \bar\rho_k)$. Though slightly unfamiliar, the  explicit
and to some extent compact expressions for the eigenvalues of $H$
and $R$ may prove very useful in extracting the spectrum and some
couplings. Equivalent but less explicit formulae have been found in \cite{lars}.
As an illustration of our procedure, we have computed
all the magnetic shifts relevant to determine the spectrum of the
${\cal N}=1$ supersymmetric AM model constructed by Antoniadis and
Maillard \cite{am}. The not so inspiring results are gathered in
the Appendix.

As usual multiplicities, associated to the degeneracy of the
Landau levels, are counted by the relevant index theorem. In six
dimensions it is given by the third Chern number $c_2(q_a {\cal
F}_a + q_b {\cal F}_b)$ of the abelian gauge bundle \bea I_{ab}
&=& - {1 \over 3! (2\pi)^3} \int_{{\Sigma}^6} Tr_{(N_a,N_b)}[(q_a
{\cal F}_a + q_b {\cal F}_b) ^3] \nonumber \\
&=& - {W_a W_b \over 3! (2\pi)^3} \int_{{\bf T}^6}
Tr_{(N_a,N_b)}[(q_a {F}_a + q_b {F}_b)^3 ] \quad .
\label{indsex}\eea For ``parallel" magnetic fields in factorized
rectangular two-tori ${\bf T}^6 = {\bf T}^2_{(1)}\times {\bf
T}^2_{(2)}\times {\bf T}^2_{(3)}$, $W_{a/b}=\prod_{i=1}^3
n^{(i)}_{a/b}$ and (\ref{indsex}) reduces to the simpler
expression \be I_{ab} = {W_a W_b \over (2\pi)^3} \prod_{i=1}^3(q_a
h^{(i)}_a + q_b h^{(i)}_b) = \prod_{i=1}^3 (q_a m^{(i)}_a
n^{(i)}_b + q_b m^{(i)}_b n^{(i)}_a) \ee upon imposing Dirac
quantization condition on each subtorus and for each stack of
branes. When $I_{ab}=0$ one has no chiral asymmetry that may
sometimes signal the presence of one or more unmagnetized
directions. This is the case \eg for the AM model described in the
appendix. One can define a reduced index $I^\perp_{ab}$ in the
subspace orthogonal to the fixed tori that coun their number.
Clearly in such cases, dycharged strings can carry generalized KK
momenta in the unmagnetized directions.

We then turn to address the issue of moduli stabilization that
relies on the potential for the closed string moduli generated by
the abelian and thus mutually commuting (in the Chan-Paton sense!)
magnetic fluxes. The Born-Infeld action, compactly expressed in
terms of the skew eigenvalues of $H_{\bf 6}$ as in (\ref{h67}),
can be rewritten as \be
 \sqrt{\prod_i (1 + h_i^2)} = \sqrt{(1 - h_1 h_2 - h_2 h_3 - h_3 h_1)^2 +
(h_1 h_2 h_3  - h_1 - h_2 - h_3)^2} \ee up to an overall factor of
$\sqrt{\det({\cal G})}$. Expressing $h_k $ in terms of the
available $SO(6)$ invariants, \ie $tr(H^2)$, $tr(H^4)$, and
$tr(H^6)$ or, equivalently, $\varepsilon HHH = - tr(H^6)/6 +
tr(H^2)tr(H^4)/8 - tr(H^2)^3/48$, requires solving a complete
cubic equation. For $F_{(2,0)}=0$, we have already solved the
problem, since $H_{\bf 6} = H_{\bf 3} + H_{{\bf 3}^*}$ and $h_k
=\pm \mu_k$. In turn, $\mu_k$ are given in terms of the only
available $U(3)$ invariants $h_0$, $|h|^2$, and $dhhh$. Combining
with the volume factor yields \cite{mmms} \be {|W|\over 3!}\sqrt{
\left(\int J\wedge J \wedge J -3\int J\wedge F \wedge F\right)^2 +
\left(\int F\wedge F \wedge F - 3\int J\wedge J \wedge F\right)^2}
\ee

{}{}{}{} In a supersymmetric configuration the last term is absent
due to (\ref{stabil}) or equivalently to the condition \be h_1 h_2
h_3  = h_1 + h_2 + h_3 \ee which can be interpreted in  geometric
terms as saying that $\beta_i=\arctan(h_i)$ are the three internal
angles of a triangle. As a result the effective tension, which is
positive as a whole may
 have negative contribution on some subtori. This is exactly what is needed to
cancel R-R tadpole while preserving susy and without introducing
lower dimensional $\Omega$-planes \cite{antidsusy,am,dt}. In a
sense magnetized branes behave as bound states of lower
dimensional branes and anti-branes that can carry (partial)
tensions and charges of different signs.    The very consistency
of configurations of this kind requires stabilizing the complex
structure and K\"ahler moduli at special values depending on the
magnetic fluxes. The stabilization mechanism proposed in \cite{am}
for the closed string moduli is rather simple and effective.
Imposing the holomorphy condition $F^a_{(2,0)}=0$ for a large
enough number of stacks with oblique magnetic fluxes fixes the
complex structure moduli.  Imposing the stability condition
$F^a_{(1,1)}\wedge F^a_{(1,1)} \wedge F^a_{(1,1)} = 3 J\wedge J
\wedge F^a_{(1,1)}$ then fixes the K\"ahler moduli. This mechanism
is particularly clear for the NS-NS moduli $G_{ij}$, both complex
structure $G_{IJ}$ and  K\"ahler deformations $G_{I\bar{J}}$,
which appear in the Born-Infeld action and for the R-R scalars
$C_{I\bar{J}}$ associated to K\"ahler deformations which can
combine with the anomalous magnetic $U(1)$ gauge fields and become
massive \cite{am,gss,u1an,mbjfm}. However the R-R scalars
$C_{I{J}}$ associated to complex structure deformations cannot
directly mix with gauge fields whose internal field strenghts are
of complex type (1,1) at the supersymmetric point. The presence of
internal magnetic fluxes, however, allows all R-R fields to mix
with one another and with NS-NS fields even at zero spacetime
momentum. In order to show this one has to generalize the results
of \cite{gm,hk}, where closed string scattering amplitudes on
Dp-branes were computed, to the case of our interest, \ie
D9-branes with internal magnetic fields. Setting $\ap = 1$ for
simplicity, the relevant amplitudes are all of the form  \be {\cal
A}(p_1,\xi_1; p_2,\xi_2) = -{i\over 2}\kappa \tau_p {\Gamma(s)
\Gamma(t) \over \Gamma(1 + t + s)} {\cal K}(p_1,\xi_1; p_2,\xi_2)
\ee where $s=p_1 D p_1$, respectively $t=(p_1+p_2)^2$, denote the
kinematic invariants in the open, respectively closed, string
exchange channels, while $\xi_1,\xi_2$ indicate the polarizations
of the external closed string states. The boundary reflection
conditions are coded in the matrix $D$. For a Dp-brane with $F=0$,
one has $D_{\mu\nu}= \eta_{\mu\nu}$, $D_{\mu i}=D_{j \nu } = 0$,
$D_{ij} = - \delta_{ij}$. Momentum conservation along the
longitudinal $p+1$ directions is understood.

We are interested in adapting the original amplitudes to the case
$F \neq 0$ in the internal directions, so that $D (F)= \identmat
\oplus R(F)$ and $M (F)=\identmat \otimes \Lambda(F)$, where
$\Lambda(F)$ is the spinor representation of $R(F)$. We focus on
the two-point amplitude mixing an internal metric fluctuation,
$\phi_{ij} = \phi_{ji}$, with an internal R-R field, $\chi_{ij} =
- \chi_{ji}$. Imposing spacetime momentum conservation only at the
very end and slightly relaxing the mass-shell condition, by
allowing some small non conserved internal momentum / winding,
yields
\be {\cal K}^{^{R-R/NS-NS}}
 = \chi^{(1)}_{ij}\phi^{(2)}_{mn} [i{s\over 2\sqrt{2}} (s+t)
R^n{}_{k}\langle\gamma^{ij} \Lambda \gamma^{k}\gamma^{m}\rangle +
i{t\over (2)\sqrt{2}} t \langle \gamma^{ij} \Lambda \rangle
R^{mn}] \ee where $\langle ...\rangle$ stands for a trace over
spinor indices. The small momentum expansion of the Veneziano-like
amplitude yields \be {\Gamma(s) \Gamma(t) \over \Gamma(1 + t + s)}
= {1\over st} - {\pi^2\over 6} + {\cal O}(p^4) \ee that exposes
massless poles in the $s$ and $t$ channel.

A similar analysis for the scattering of massless closed strings
on $\Omega$-planes gives a result of the form \be {\cal
A}(p_1,\xi_1; p_2,\xi_2) = -{i\over 2t(s+t)}\kappa \tau_\Omega
F(1-s,t; t+1; -1) {\cal K}_{\Omega}(p_1,\xi_1; p_2,\xi_2) \ee
where $ F(1-s,t; t+1; -1)= t \int_0^1 dy y^{t-1} (1 + y)^{s-1}$ is
a hypergeometric function.

The massless $t$ channel poles of the various amplitudes
correspond to closed string massless tadpoles in the NS-NS and R-R
sectors. They cancel in an anomaly free supersymmetric backgrounds
\cite{mbas,mbthe,chiralas,gep,bmp,schel}, where the scalar fields
extremize the potential, after summing the contributions of
(magnetized) D-branes and $\Omega$-planes .

The $s$ channel poles correspond to the exchange of massless open
string states that mix with the closed string states. This mixing
receives no contribution from the projective plane that cannot
accomodate open string channels. Indeed $F(1-s,t; t+1; -1)$ is
finite for $s=0$. As argued in \cite{am} this open-closed string
mixing is responsible for lifting part of the moduli, \ie the R-R
axions $\chi_{I\bar{J}}$ of type (1,1) partecipating in the
generalized GS mechanism of cancellation of the anomalous $U(1)$'s
that schematically reads \be\langle F^3\rangle_{1-loop} - \langle
F \beta\rangle_{1/2-loop} \langle \beta \chi\rangle_{tree}
\langle\chi FF\rangle_{1/2-loop} =0 \label{gss}\ee where
$\beta_{\mu\nu KL\bar{M}\bar{N}}$ is the R-R two-form dual (in
4-d) to the axionic R-R scalars $\chi_{I\bar{J}}$.

Even after subtracting the open string massless poles, there is a
finite remnant that leads to R-R / NS-NS mixing of the schematic
form \be {\cal L}_{mix} = \chi_{I\bar{J}} \phi_{K\bar{L}}
\langle\gamma^{I\bar{J}} \Lambda \gamma^K \gamma_M\rangle
R^{M\bar{L}} \nonumber + \chi_{IJ} \phi_{\bar{K}\bar{L}}
\langle\gamma^{IJ} \Lambda \gamma^{\bar{K}} \gamma_M\rangle R^{M
\bar{L}} \ee These include terms responsible for the mixing of the
(2,0)-type R-R fields with the NS-NS (2,0) type metric
fluctuations and for their mass generation. At first sight this
looks puzzling\footnote{M.B. would like to thank C.~Angelantonj,
S.~Ferrara and J.~F.~Morales Morera for a very enlightening
discussion on this topic.} since a non derivative coupling of R-R
scalars seems to violate invariance under their axionic shift
symmetries. Moreover the standard form of the WZ couplings seems
to prevent such mixings. However one should not forget that the
R-R 6-form $C_6$ that appears in $S_{WZ}$ is not an elementary
field but rather a composite object arising from $C_2$ after
Poincar\'e duality. This in turn involves the metric as required
for the mixing. Another way to see this is simply solving for
$C_6$ and working with the modified 3-form field strength for
$C_2$, \ie $H_3 = dC_2 + e^\phi \omega^{CS}_3$. The beating term
$e^{-\phi} dC_2\wedge *\omega^{CS}_3$ in the expansion of the
kinetic term for $C_2$ arises at the correct order (disk) and once
integrated by parts can produce the desired mixing with the NS-NS
metric fluctuations.

A detailed resolution of the mixing requires dealing with subtle
relative normalization issues among the various amplitudes  and is
beyond the scope of the present investigation. In specific models,
such as the AM model, one can accomplish the task with some effort
and motivation. Suffice it to say that the resulting masses are
highly non-trivial functions of the fluxes but, being
supersymmetric, should satisfy sum rules such as $Str({\cal
M}^2)=0$. In principle this could be checked by computing
two-point amplitudes of closed string fermions on the disk and
projective plane.

Before concluding this section let us briefly mention the role of
open string moduli (Wilson lines) \cite{bpstor} and of additional
fluxes associated to metric torsion (Scherk-Schwarz deformations
\cite{ss,sshet,ssopen}) and to internal 3-form fluxes
\cite{3formflux}. By close analogy with the heterotic string on
twisted tori \cite{km}, one can easily conclude that part of the
open string Wilson line moduli are eaten by the six NS-NS
graviphotons $G_{\mu i}$ very much as part of the R-R axions are
eaten by the `anomalous' $U(1)$ vector bosons $A_{\mu a}$ in the
open string spectrum. Adding metric torsion contributes extra mass
terms for the six R-R graviphotons $B_{\mu i}$ in a perturbative
setting. Indeed Scherk-Schwarz deformations of type I
compactifications have been under active investigation for some
time \cite{sshet,ssopen}. In order to stabilize the dilaton and
the R-R axion it seems however necessary to turn on internal R-R
3-form fluxes\footnote{M.~B. would like to thank I.~Antoniadis for
a discussion on this point.}. This would spoil the nice
perturbative framework of the analysis performed here. In any case
it would be very important to study the conditions for moduli
stabilization and unbroken susy in the presence of combined
constant fluxes from the three sectors of type I strings: magnetic
fields ($m^a_{ij}$), NS-NS torsion \`a la Scherk-Schwarz
($\gamma^k_{ij}$) and R-R fluxes ($\beta_{ijk}$). We cannot help
writing down the resulting non-abelian algebra satisfied by the
generators of internal translations $P_i$, Cartan subalgebra of
the Chan-Paton group $T_a$ and axionic shifts
 $W^i$ \cite{km}
 \bea &&[P_i, P_j] = 2i m^a_{ij} T_a + 2i
\gamma^k_{ij}P_k - 3i \beta_{ijk} W^k \nonumber \\
&&[P_i, T^a] = - 2i m^a_{ij}W^j T_a \qquad, \qquad [P_i, W^k] = -
2i
\gamma^k_{ij} W^j \\
&&[W^i, W^k]=0 \quad , \quad [T_a, W^k]=0 \quad , \quad [T_a,
T_b]=0 \quad .\eea The structure constants are constrained by
Jacobi identities \bea &&\gamma^l_{n[i}\gamma^n_{kj]} = 0 \quad , \quad \gamma^k_{kj}=0 \\
&& m_{a[ij}m^a_{kl]} = 3 \beta_{n[ij}\gamma^n_{kl]} \quad , \quad
m^a_{n[k}\gamma^n_{ij]} = 0 \nonumber \eea which admit a geometric
interpretation in terms of Bianchi identities for the internal
p-form fields.

 {} {} {{}} {{{}}}
\section{Conclusions and perspectives}
\label{conclus}

The results presented above should have clarified several features
of the dynamics of open strings coupled to oblique magnetic
fields. In turn these can give a host of new possibilities for
symmetry breaking, moduli stabilization and chiral asymmetry even
in toroidal compactifications while preserving some residual
supersymmetry. Clearly these notes were aimed a general analysis
rather then the construction of specific models\footnote{Models with (oblique) magnetic fluxes are subject to subtle consistency conditions in order to cancel K-theory RR charges. We thank F. Marchesano for pointing this out to us.} where all or some
of the above effects are present. Some explicit models have been
very recently discussed in \cite{Marchesano:2004yq,Marchesano:2004xz,dt}, 
where a puzzle first raised
in \cite{aseri} has been successfully solved but only ``parallel''
magnetic field have been considered. More phenomenological aspects
of oblique internal magnetic fields are under investigation
\cite{prep}.

Before concluding, let us briefly discuss toroidal orbifolds and
models which are based on rational conformal field theories
\cite{mbas,mbthe,bmp}, such as Gepner models \cite{gep,schel}. As
shown in \cite{mbys} for the case of the NS penta-brane, governed
by the product of an $SU(2)_k$ WZW model and a non-compact free
boson with a background charge $Q_k$, turning on magnetic fields
corresponds to changing the phase of some boundary reflection
coefficients in the transverse closed string channel. This in turn
generates ``twisted'' representations in the direct open string
channel, whose modes are shifted by an {\it a priori} rather
arbitrary amount. In the case studied in \cite{mbys} supersymmetry
is completely broken by the introduction of magnetic fluxes but in
more general cases where the chiral algebra on the worldsheet
admits a spectral flow it should be possible to tune them in such
a way as to preserve some supersymmetry \cite{mmms,am,fieri}.

As pointed out above, the non-linearity of the abelian Born-Infeld
action allow for more general possibilities of non BPS brane
configurations that accomodate magnetic fields stisfying a
generalized non-linear (anti) self-duality condition of the type
$H_r = 1/H_s$ where $r$ and $s$ label two (1,1) forms. This kind
of branes may offer further opportunities for phenomenologically
viable type I models or their T-dual intersecting brane models
\cite{bcms,bcls}.

It is important to stress that contrary to the case of combined
closed  string NS-NS and R-R fluxes \cite{kst,3formflux}, abelian
magnetic fields coupled to the ends of open strings are completely
under control in string perturbation theory. One can
straightforwardly compute the spectrum and interactions. In the
present notes we have focussed on how to derive the spectrum, in
terms of magnetic shifts, multiplicities and rescalings of the
internal KK momenta, but it is rather easy to generalize our
analyis and extract some relevant low-energy couplings in addition
to the scalar potential. Very much as in intersecting brane models
which are T-dual to type I compactifications with ``parallel"
magnetic fluxes \cite{gauge}, tree-level gauge couplings are
simply given by \be {1\over g_a^2} = |W_a| e^{-\phi}
\sqrt{\det(G+F_a)} \ee that further simplify at supersymmetric
extrema. Notice the role of the wrapping factors
$|W_a|=\det(W_a)^i_\alpha$ in the relative rescaling of the gauge
couplings wrt to one another. Due to R-R tadpole cancellation
$|W_a|$ are also important in reducing the rank of the gauge
group. The one-loop running can be determined turning on (small)
magnetic fields along the non-compact spacetime directions as in
\cite{gauge}. The role of the $B$ field has been investigated by \cite{Mihailescu:2000dn}

Another important aspect that has been investigated in \cite{am}
is the possibility of accomodating some large extra dimensions
\cite{aahdd}. We hope to get back to this point in \cite{prep}
together with a detailed analysis of Yukawa couplings for models
with oblique internal magnetic fields extending the analysis of
the ``parallel'' case in \cite{yukawa,abesch}.

\section*{Acknowledgements}
We would like to thank P.~Anastasopoulos, C.~Angelantonj,
I.~Antoniadis, S.~Ferrara, F.~Fucito, M.~Larosa, A.~Lionetto,
J.~F.~Morales Morera, M.~Nicolosi, G.~Pradisi, M.~Prisco,
R.~Russo, A.~Sagnotti, and Ya.~Stanev for useful discussions. Our
special thanks go to C.~Angelantonj, G.~Pradisi, A.~Sagnotti for
valuable comments on the manuscript. M.~B. would like to thank the
CERN Theory Division for its kind hospitality while this work was
being completed. This work was supported in part by INFN, by the
MIUR-COFIN contract 2003-023852, by the EU contracts
MRTN-CT-2004-503369 and MRTN-CT-2004-512194, by the INTAS contract
03-516346 and by the NATO grant PST.CLG.978785.

\newpage
\section*{Appendix: the AM model}
In this appendix we display the rotation matrices $R_a$ and list
the eigenvalues of $R_a^{q_a}R_b^{q_b}$ for the AM toroidal  model
constructed by Antoniadis and Maillard \cite{am}. It consists of 9
stacks of magnetized D9-branes and one stack of unmagnetized ones.
An explicit realization of moduli stabilization through
``tangled'' $U(1)$ fluxes is shown. In the following table, we
list the number of branes and the numerical value of the internal
magnetic fields at the stable supersymmetric point.

\begin{tabular}{|r|c|l|}
\hline
Stacks & magnetic and wrapping numbers & magnetic fields\\
\hline
$N_1 = 1$ &  $(m_{x_1 y_2}, n_{x_1 y_2}) = (1,-1)$ &$H_{x_1 y_2}= -\sqrt{2}$\\
 &   $(m_{x_2 y_1},n_{x_2 y_1}) = (1,- 1)$ & $H_{x_2 y_1}= -\sqrt{2}$\\
&    $(m_{x_3 y_3},n_{x_3 y_3}) = (0,-1)$ & $ H_{x_3 y_3}= 0$\\
\hline
$N_2 = 1$ &  $(m_{x_1 y_3}, n_{x_1 y_3}) = (1,-1)$ &$H_{x_1 y_3}= -2$\\
 &   $(m_{x_3 y_1},n_{x_3 y_1}) = (1,-1)$ & $H_{x_3 y_1}= -2$\\
&    $(m_{x_2 y_2},n_{x_2 y_2}) = (0,-1)$ & $ H_{x_2 y_2}= 0$\\
\hline
$N_3 = 1$ &  $(m_{x_1 x_2}, n_{x_1 x_2}) = (1,-1)$ &$H_{x_1 x_2}= -\sqrt{2}$\\
 &   $(m_{y_1 y_2},n_{y_1 y_2}) = (1,-1)$ & $H_{y_1 y_2}= -\sqrt{2}$\\
&    $(m_{x_3 y_3},n_{x_3 y_3}) = (0,-1)$ & $ H_{x_3 y_3}= 0$\\
\hline
$N_4 = 2$ &  $(m_{x_1 y_1}, n_{x_1 y_1}) = (0,-1)$ &$H_{x_1 y_1}= 0$\\
 &   $(m_{x_2 x_3},n_{x_2 x_3}) = (1,-1)$ & $H_{x_2 x_3}= -1 $\\
&    $(m_{y_2 y_3},n_{y_2 y_3}) = (1,-1)$ & $ H_{y_2 y_3}= -1$\\
\hline
$N_5 = 1$ &  $(m_{x_1 x_3}, n_{x_1 x_3}) = (1,-1)$ &$H_{x_1 x_3}= -2$\\
 &   $(m_{x_2 y_2},n_{x_2 y_2}) = (0,-1)$ & $H_{x_2 y_2}= 0 $\\
&    $(m_{y_1 y_3},n_{y_1 y_3}) = (1,-1)$ & $ H_{y_1 y_3}= -2$\\
\hline
$N_6 = 2$ &  $(m_{x_1 y_1}, n_{x_1 y_1}) = (0,-1)$ &$H_{x_1 y_1}= 0$\\
 &   $(m_{x_2 y_3},n_{x_2 y_3}) = (1,-1)$ & $H_{x_2 y_3}= -1 $\\
&    $(m_{x_3 y_2},n_{x_3 y_2}) = (1,-1)$ & $ H_{x_3 y_2}= -1$\\
\hline $N_7 = 2$ &  $(m_{x_1 y_1}, n_{x_1 y_1}) = (-2,1)$ &$H_{x_1
y_1}=
-\frac{1}{\sqrt{2}}$\\
& $(m_{x_2 y_2}, n_{x_2 y_2}) = (0,1)$ &
$H_{x_2 y_2}= 0$\\
& $(m_{x_3 y_3},n_{x_3 y_3}) = (1,1)$ &
$ H_{x_3 y_3}= \frac{1}{\sqrt{2}}$\\
\hline $N_8 = 2$ &  $(m_{x_1 y_1}, n_{x_1 y_1}) = (-1,1)$ &$H_{x_1
y_1}=
-\frac{1}{2\sqrt{2}}$\\
 &   $(m_{x_2 y_2},n_{x_2 y_2}) = (1,1)$ & $H_{x_2 y_2}= \sqrt{2} $\\
&    $(m_{x_3 y_3},n_{x_3 y_3}) = (-1,1)$ & $ H_{x_3 y_3}=
-\frac{1}{\sqrt{2}}$\\
\hline
$N_9 = 1$ &  $(m_{x_1 y_1}, n_{x_1 y_1}) = (0,1)$ &$H_{x_1 y_1}=0 $\\
 &   $(m_{x_2 y_2},n_{x_2 y_2}) = (-1,1)$ & $H_{x_2 y_2}= -\sqrt{2} $\\
&    $(m_{x_3 y_3},n_{x_3 y_3}) = (2,1)$ & $ H_{x_3 y_3}= \sqrt{2}$\\
\hline
$N_0 = 3$ &  $(m_{ij}, n_{ij}) = (0,\pm 1)$ &$H_{ij}=0 $\\
\hline
\end{tabular}

Notice that all wrapping numbers are equal to $\pm 1$.

The first six stacks of magnetized branes, each preserving ${\cal
N}=2$ susy, share a common ${\cal N}=1$ susy. The supersymmetry
conditions in (\ref{susycond})
 fix the nine deformations (6 NS-NS, 3 R-R) of the complex structure to
be diagonal and purely imaginary, $\tau_{IJ} = i \delta_{IJ}$. In
addition there are three stacks of magnetized branes, two of which
preserve ${\cal N}=2$ susy, while the remaining one preserves the
${\cal N}=1$ susy shared by all the branes. The stability
conditions, trivially satisfied by  the first six stacks, fix the
K\"ahler structure when imposed on the latter three stacks.

The resulting K\"ahler form reads $J = J_{(1)} + J_{(2)} + J_{(3)}$ with \bea
J_{(1)} & = & 4\pi^2 2 \sqrt{2}\ap \frac{dz_1 \wedge d\bar{z_1}}{2i}\\
J_{(2)} & = & \frac{4\pi^2}{\sqrt{2}}\ap\frac{dz_2 \wedge d\bar{z_2}}{2i}\\
J_{(3)} & = & 4\pi^2 \sqrt{2}\ap \frac{dz_3 \wedge d\bar{z_3}}{2i}
\eea \ie radii $r_1 = r_2 = 2^{3/4} \sqrt{\ap}, r_3 = r_4 =
2^{-1/4} \sqrt{\ap}, r_5 = r_6 = 2^{1/4} \sqrt{\ap}$. With the
above supersymmetric choice of fluxes, we computed the ortogonal $6\times 6$
matrix, $R_{a}$, for each stack of branes. They read
\bea R_1^{\pm 1} =
\left(
\begin{array}{ccc} - {1\over 3}{\identmat} & \mp {2\sqrt{2}\over
3}i\sigma_2 & 0 \\ \mp
{2\sqrt{2}\over 3} i\sigma_2 & -{1\over 3} {\identmat} & 0\\
0 & 0 & {\identmat} \ea \right) \quad R_3^{\pm 1} = \left(
\begin{array}{ccc} - {1\over
3}{\identmat} & \mp {2\sqrt{2}\over 3}{\identmat} & 0 \\ \pm
{2\sqrt{2}\over 3} {\identmat} & -{1\over 3} {\identmat} & 0\\
0 & 0 & {\identmat} \ea \right)   \eea

\bea R_2^{\pm 1} = \left(
\begin{array}{ccc} - {3\over 5}{\identmat} & 0 & \pm
{4\over 5}i\sigma_2 \\
0 & {\identmat} & 0\\
\pm {4\over 5}i\sigma_2  & 0 & - {3\over 5}{\identmat}\ea\right)
\quad R_5^{\pm 1} = \left(
\begin{array}{ccc} - {3\over 5}{\identmat} & 0 & \pm
{4\over 5} {\identmat}\\
0 & {\identmat} & 0\\
\mp {4\over 5}{\identmat}  & 0 & - {3\over 5}{\identmat}\ea
\right) \eea

\bea R_4^{\pm 1} = \left(
\begin{array}{ccc} {\identmat} & 0 & 0 \\
0 & 0 & \mp{\identmat} \\
0  & \pm{\identmat} & 0 \ea\right) \quad R_6^{\pm 1} = \left(
\begin{array}{ccc} {\identmat} & 0 & 0 \\
0 & 0 & \pm i\sigma_2 \\
0  & \pm i\sigma_2 & 0 \ea\right)  \eea

\bea R_7^{\pm 1} = \left(
\begin{array}{ccc} {1\over 3}{\identmat}  \pm {2\sqrt{2}\over 3} i\sigma_2 & 0 & 0 \\
0 & {\identmat}  & 0 \\
0  & 0 & {1\over 3} {\identmat}  \mp {2\sqrt{2}\over 3}
i\sigma_2\ea\right) \eea

\bea R_8^{\pm 1} = \left(
\begin{array}{ccc} {7\over 9}{\identmat}  \pm {4\sqrt{2}\over 9} i\sigma_2& 0 & 0 \\
0 & -{1\over 3} {\identmat}  \mp {2\sqrt{2}\over 3} i\sigma_2  & 0 \\
0  & 0& {1\over 3} {\identmat}  \pm {2\sqrt{2}\over 3}
i\sigma_2\ea\right) \eea

\bea R_9^{\pm 1} = \left(
\begin{array}{ccc} {\identmat}  & 0 & 0 \\
0 & -{1\over 3} {\identmat}  \pm {2\sqrt{2}\over 3} i\sigma_2  & 0 \\
0  & 0& -{1\over 3} {\identmat} \mp {2\sqrt{2}\over 3}
i\sigma_2\ea\right) \eea

Switching to complex coordinates $ z^{i}=x^{i}+iy^{i} $
the above matrices become block diagonal \bea R_a^{\pm 1} = \left(
\begin{array}{cc} U_{a}^{\pm 1} & 0\\
0 & U_{a}^{ *\,\pm 1}\ea\right) \eea where the sub-blocks $U_{a}$
are $3\times 3$ complex matrices belonging to $SU(3)\subset
SU(4)$. They read

\bea U_1^{\pm 1} = \left(\begin{array}{ccc} - {1\over 3} & \mp
{2\sqrt{2}\over 3}i & 0\\ \mp
{2\sqrt{2}\over 3} i & -{1\over 3}  & 0\\
0 & 0 & 1 \ea \right) \quad U_3^{\pm 1} = \left(\begin{array}{ccc}
- {1\over 3}& \pm {2\sqrt{2}\over 3} & 0\\ \mp
{2\sqrt{2}\over 3}  & -{1\over 3} & 0\\
0 & 0 & {1}\ea\right)   \eea

\bea U_2^{\pm 1} = \left(\begin{array}{ccc} - {3\over 5}& 0 & \mp
{4\over 5}i \\
0 & 1 & 0\\
\mp {4\over 5}i & 0 & - {3\over 5}\ea\right) \quad U_5^{\pm 1} =
\left(\begin{array}{ccc}- {3\over 5} & 0 & \pm
{4\over 5}  \\
0 & 1 & 0\\
\mp {4\over 5}  & 0 & - {3\over 5}\ea\right)  \eea

\bea U_4^{\pm 1} = \left(\begin{array}{ccc} 1 & 0 & 0 \\
0 & 0 & \pm 1 \\
0  & \mp 1 & 0\ea\right)\quad U_6^{\pm 1} =
\left(\begin{array}{ccc} 1 & 0 & 0
\\
0 & 0 &\mp i \\
0  & \mp i  & 0\ea\right) \eea

\bea U_7^{\pm 1} = \left(\begin{array}{ccc} {1\over 3} \mp
{2\sqrt{2}\over 3}
i & 0 & 0 \\
0 & 1 & 0 \\
0  & 0 & {1\over 3} \pm {2\sqrt{2}\over 3} i\ea\right) \eea \bea
U_8^{\pm 1} = \left(\begin{array}{ccc} {7\over 9}\mp
{4\sqrt{2}\over 9} i& 0 &
0 \\
0 & -{1\over 3} \pm {2\sqrt{2}\over 3} i  & 0 \\
0  & 0& {1\over 3} \mp {2\sqrt{2}\over 3} i\ea\right) \eea

\bea
U_9^{\pm 1} = \left(\begin{array}{ccc} 1 & 0 & 0 \\
0 & -{1\over 3} \mp {2\sqrt{2}\over 3} i  & 0 \\
0  & 0& -{1\over 3} \pm {2\sqrt{2}\over 3} i\ea\right) \eea

Except for the stacks labelled by $a=8$, all
the submatrices have, at least, one unit eigenvalue
(in fact three for $a=0$). The associated eigenvector defines a
(complex) unmagnetized direction with standard KK momenta.
The other two eigenvalues are complex conjugate.

As usual the open string states can be grouped into three main
sectors: neutral, singly and doubly charged ones. With  an abuse
of notation we will often call magnetic shifts the angles $\beta =
\pi \epsilon$.

The uncharged sector comprise open strings stretching either
between two unmagnetized branes and carrying standard KK momenta
or strings starting and ending on the same stack of  magnetized
branes ($q_a=-q_b$)and carrying reshuffled KK momenta and
windings.

In the singly charged sector, one endpoint of the string  is on
the $N_0$ stack, the other is on one of the stack of magnetized
branes, $N_a$, or their images under $ \Omega $, $ N_{a}^{*}$. The
frequencies of the oscillator modes are shifted by\be
{\beta^{(k)}_{0a}} =\frac{\arg(\rho^{(k)}_a)}{2\pi} \ee Open
strings in this sector appear with a multiplicity due to the
degeneracy of the Landau levels. For $a=8$, this is given by the
product of the magnetic numbers $\prod_{I} m^I_a$, in all other
cases it is reduced to $\prod_{I\neq u} m^I_a $ for the presence
of some unmagnetized plane,  generically indicated by $u$.

Charged strings, starting on  magnetized stack and arriving at its
images are doubly charged ($q_a=q_b=\,\pm 1$). The frequencies
shift is \be {2\beta^{(k)}_{aa}} ={arg(\rho^{(k)}_a)\over \pi}\ee
and the multiplicities are $\prod_{I}(2 m^I_8)$ for $a=8$ or
$\prod_{I\neq u} (2m^I_a)$ for all $a \neq 8 $.

The situation becomes more involved when strings, starting on
magnetized stack of branes (say, $ N_a $) and ending on a different
stack (labelled by  $ N_b^{*} $, or its image ),  are taken into
account. In all, there are 36 distinct [a,b] sectors.

In the three sectors [1,3],[2,5],[4,6] diagonalization of
 $R_a^{q_a} R_b^{q_b}$ yields one $\rho^u_{ab}=1$ and two complex conjugate pairs determined by
$$ \tan(2\beta^I_{ab}) = \pm \sqrt{t_{a}^2 +
t_{b}^2 + t_{a}^2t_{b}^2} \quad  \mbox{ for }I\neq u$$ where
$t_{a/b}=\tan(\beta^I_{a/b})$.

In the 3 sectors [7,8],[7,9],[8,9],  due to diagonal form of the
submatrices $U_{i}^{q_a}$, the magnetic shifts are simply
$$
\beta^I_{ab}=\pm\beta^I_{a}\pm\beta^I_{b}
$$
and there is no unmagnetized direction.

The twelve sectors [13]-[25], [25]-[46] and [46]-[13] all have one
unit eigenvalue (one complex unmagnetized direction/plane) and a
complex conjugate pair, $\rho^{\pm}_{ab}$. More explicitly one finds
\bea
&&\rho^{\pm}_{[1,3]-[2,5]}=-\frac{1}{15}(13\mp
2\sqrt{14}) \\
&&\rho^{\pm}_{[2,5]-[4,6]}=-\frac{4}{5}\pm
\frac{3 i}{5} \\
&&\rho^{\pm}_{[4,6]-[1,3]}=-\frac{2}{3}\pm
\frac{i\sqrt{5}}{3} \quad . \eea

The remaining 18 sectors given by [789]-[123456] can have common
unmagnetized planes and frequencies given by \bea
\beta^I_{Lab} & = & \beta^I_{La}\nonumber\\
\quad\tan(2\beta^I_{Rab}) & = & \pm \sqrt{t_{aR}^2 + t_{bR}^2 +
t_{aR}^2t_{baR}^2}\eea
 for $I\neq u$. The results are listed in the following tables.

\begin{table}
\begin{tabular}{|r|c|l|}
\hline
sectors & eigenvalues & unmagnetized directions\\
\hline
$U_{4}^{q_{a}}U_{1}^{q_{b}}$ &$ \rho= 1 \mbox{ and }
\rho^{\pm}=(-2\pm
i\sqrt{5})/ 3$ & $(-\frac{iq_{a}q_{b}}{\sqrt{2}},q_{a},1)$\\
\hline
$U_{4}^{q_{a}}U_{3}^{q_{b}}$ & $ \rho= 1 \mbox{and }
\rho^{\pm}=(-2\pm
i\sqrt{5})/ 3$ & $(\frac{q_{a}q_{b}}{\sqrt{2}},q_{a},1)$\\
\hline
$U_{6}^{q_{a}}U_{1}^{q_{b}}$ &$ \rho= 1 \mbox{and }
\rho^{\pm}=(-2\pm
i\sqrt{5})/ 3$ &$( -\frac{q_{a}q_{b}}{\sqrt{2}},-iq_{a},1)$ \\
\hline
$U_{6}^{q_{a}}U_{3}^{q_{b}} $&$ \rho= 1 \mbox{and }
\rho^{\pm}=(-2\pm
i\sqrt{5})/ 3
$ & $( -\frac{iq_{a}q_{b}}{\sqrt{2}},-q_{a},1)$\\
\hline
$U_{2}^{q_{a}}U_{4}^{q_{b}}$&  $ \rho= 1 \mbox{ and }
\rho^{\pm} = (-4\pm i\,3)/5$ & $\frac{iq_{a}}{2},q_{b},1)$\\
\hline
$U_{2}^{q_{a}}U_{6}^{q_{b}}$ & $ \rho= 1 \mbox{ and }
\rho^{\pm} = (-4\pm i\,3)/5$ &$(\frac{iq_{a}}{2},-iq_{b},1)$\\
\hline
$U_{5}^{q_{a}}U_{4}^{q_{b}}$ & $ \rho= 1 \mbox{ and }
\rho^{\pm} = (-4\pm i\,3)/5 $ & $ (-\frac{q_{a}}{2},q_{b},1)$\\
\hline
$U_{5}^{q_{a}}U_{6}^{q_{b}}$ & $ \rho= 1 \mbox{ and }
\rho^{\pm} = (-4\pm i\,3)/5
$ & $(-\frac{q_{a}}{2},-iq_{b},1)$\\
\hline
$U_{1}^{q_{a}}U_{2}^{q_{b}}$  & $\rho= 1 \mbox{ and }
\rho^{\pm} = (-13\pm
\sqrt{56})/15$ & $(2iq_{b},-\sqrt{2}q_{a}q_{b},1)$\\
\hline
$U_{1}^{q_{a}}U_{5}^{q_{b}}$ &$\rho= 1 \mbox{ and }
\rho^{\pm} = (-13\pm
\sqrt{56})/15$ & $(-2q_{b},-i\sqrt{2}q_{a}q_{b},1)$\\
\hline
$U_{3}^{q_{a}}U_{2}^{q_{b}}$ & $\rho= 1 \mbox{ and }
\rho^{\pm} = (-13\pm
\sqrt{56})/15
$ &$(2iq_{b},i\sqrt{2}q_{a}q_{b},1)$\\
\hline
$ U_{3}^{q_{a}}U_{5}^{q_{b}}$ & $\rho= 1 \mbox{ and }
\rho^{\pm} =(-13\pm
\sqrt{56})/15
$ & $(-2q_{b},-\sqrt{2}q_{a}q_{b},1)$\\
\hline
\end{tabular}
\label{tab2}
\end{table}

\begin{table}
\begin{tabular}{|r|c|l|}
\hline
sectors & eigenvalues & unmagnetized directions\\
\hline
$U_{1}^{q_{a}}U_{3}^{q_{b}} $&
$\rho=1 \mbox{ and }\rho^{\pm}=\frac{1}{9}(1\pm 4i\sqrt{5})$& $(0,0,1)$\\
\hline
$U_{2}^{q_{a}}U_{5}^{q_{b}}$ & $\rho=1 \mbox{ and }\rho^{\pm}=\frac{1}{25}(9 \pm 4i\sqrt{34})$ &
$(0,1,0)$\\
\hline
$U_{4}^{q_{a}}U_{6}^{q_{b}}$ & $\rho=1\mbox{ and }  \rho^{\pm}= \pm i$
 & $(1,0,0)$\\
\hline
$U_{7}^{q_{a}}U_{8}^{q_{b}}$ &
$\rho_{1}=- (-7+16q_{a}q_{b}+2\sqrt{2}i(2q_{b}+7q_{a}))/27$ &  \\
 & $\rho_{2}= (1+8q_{a}q_{b}+2\sqrt{2}i(q_{a}-q_{b}))/9$  & no unmagnetized directions \\
& $\rho_{3}=-(1- 2i\sqrt{2}q_{b})/3$ & \\
\hline
$U_{7}^{q_{a}}U_{9}^{q_{b}}$ &$\rho_{1}=-\frac{1}{3}(1+ 2i\sqrt{2}q_{b})$ &\\
& $\rho_{2}=-(1+ 8q_{a}q_{b} +2\sqrt{2}i(q_{a}-q_{b}))/9
$&  no unmagnetized directions\\
& $\rho_{3}=\frac{1}{3}(1- 2i\sqrt{2}q_{a})$ & \\
\hline
$U_{8}^{q_{a}}U_{9}^{q_{b}}$ & $\rho_{1}=\frac{1}{9}(7-4i\sqrt{2}q_{a})$ &\\
&$\rho_{2}
=(1+ 8q_{a}q_{b} -2\sqrt{2}i(q_{a}-q_{b}))/9
 $ &
 no unmagnetized directions\\
& $\mbox{ and }
\rho_{3}=(-1+ 8q_{a}q_{b} +2\sqrt{2}i(q_{a}+q_{b}))/9$ & \\
\hline
\end{tabular}
\label{tab1}
\end{table}
\begin{table}
\begin{tabular}{|r|c|l|}%\label{tab3}
\hline
sectors & eigenvalues & unmagnetized directions\\
\hline
$U_{7}^{q_{a}}U_{1}^{q_{b}}$ &$ \rho_{1}=\frac{1}{3}(1+2i\sqrt{2}q_{a})$ &\\
&$\rho_{2,3}=\frac{1}{9}(-2 +i\sqrt{2}q_{a}\pm 5\sqrt{(-1 +2i\sqrt{2}q_{a})}$
& no unmagnetized directions\\
\hline
$U_{7}^{q_{a}}U_{2}^{q_{b}}$ &$\rho_{1}=1\mbox{ and }\rho_{2,3}=\frac{1}{15}(-3 \pm
6i\sqrt{6} $& (0,1,0)\\
\hline
$U_{7}^{q_{a}}U_{3}^{q_{b}}$ & $\rho_{1}=\frac{1}{3}(1+2i\sqrt{2}q_{a}) $ & \\
& $\rho_{2,3}=\frac{1}{9}(-2 +i\sqrt{2}q_{a} \pm 5\sqrt{(-1+2i\sqrt{2}q_{a})} $&
no unmagnetized directions\\
\hline
$U_{7}^{q_{a}}U_{4}^{q_{b}}$ &
$ \rho_{1}=\frac{1}{3}(1-2i\sqrt{2}q_{a})$ & \\
& $\rho_{2,3}=\pm\frac{\sqrt{-1-2i\sqrt{2}q_{a}}}{\sqrt{3}} $ & no unmagnetized directions\\
\hline
$U_{7}^{q_{a}}U_{5}^{q_{b}}$ & $\rho_{1}=1\mbox{ , } \rho_{2,3}=\frac{1}{15}(-3 \pm
6i\sqrt{6})$ &(0,1,0)\\
\hline
$U_{7}^{q_{a}}U_{6}^{q_{b}}$ & $ \rho_{1}=\frac{1}{3}(1-2i\sqrt{2}q_{a} $&\\
& $\rho_{2,3}=\pm\frac{\sqrt{-1-2i\sqrt{2}q_{a}}}{\sqrt{3}}$ &  no unmagnetized directions\\
\hline
$U_{8}^{q_{a}}U_{1}^{q_{b}}$ &$
\rho_{1}=\frac{1-2i\sqrt{2}q_{a}}{3}$ & \\
& $\rho_{2,3}= \frac{1}{27}(-2-i(\sqrt{2}q_{a}\pm\sqrt{241}))$ &  no unmagnetized directions\\
\hline
$U_{8}^{q_{a}}U_{2}^{q_{b}}$ &
$\rho_{1}=-\frac{1-2i\sqrt{2}q_{a}}{3}$&\\
& $\rho_{2,3}=\frac{1}{3}(-1 +i\sqrt{2}q_{a}\pm\sqrt{(2+4i\sqrt{2}q_{a})}) $ &  no unmagnetized directions\\
\hline
$U_{8}^{q_{a}}U_{3}^{q_{b}}$ &$
\rho_{1}=\frac{1-2i\sqrt{2}q_{a}}{3} $  &\\
 & $\rho_{2,3}=\frac{1}{27}(-2 -i\sqrt{2}q_{a}\pm\sqrt{(-241-482i\sqrt{2}q_{a})})$
&  no unmagnetized directions\\
\hline
$U_{8}^{q_{a}}U_{4}^{q_{b}}$ & $
\rho_{1}=\frac{1}{9}(7-4i\sqrt{2}q_{a})$ &\\
& $\rho_{2,3}=\pm\frac{1}{3}
\sqrt{(-7-4i\sqrt{2})} $ &  no unmagnetized directions\\
\hline
$U_{8}^{q_{a}}U_{5}^{q_{b}} $ & $
\rho_{1}=-\frac{1}{3}(1-2\sqrt{2}iq_{a})$  &\\ & $
\rho_{2,3}=\frac{1}{3}(-1+i\sqrt{2}\pm\sqrt{2+i4\sqrt{2}q_{a}})$ &  no unmagnetized directions\\
\hline
$U_{8}^{q_{a}}U_{6}^{q_{b}}$ &$
\rho_{1}=\frac{1}{9}(7-4i\sqrt{2}iq_{a}) $&\\ & $
\rho{2,3}=\pm\frac{1}{3}\sqrt{(-7-4i\sqrt{2}iq_{a})}$&  no unmagnetized directions\\
\hline
$U_{9}^{q_{a}}U_{1}^{q_{b}})$ &$
\rho_{1}=-\frac{1}{3}(1-2\sqrt{2}iq_{a}) $  &\\ & $
\rho_{2,3}=\frac{1}{9}(-1 +i\sqrt{2}q_{a}\pm \sqrt{26}\sqrt{(1+4\sqrt{2}iq_{a})} $&  no unmagnetized directions\\
\hline
$U_{9}^{q_{a}}U_{2}^{q_{b}}) $ &
$\rho_{1}=-\frac{1}{3}(1+2\sqrt{2}iq_{a})  $  &\\ & $
\rho_{2,3}=\frac{1}{15}(-3-3\sqrt{2}iq_{a}\pm\sqrt{6}\sqrt{(11-22i\sqrt{2}q_{a})}$ &  no unmagnetized directions\\
\hline
$U_{9}^{q_{a}}U_{3}^{q_{b}}$ &$
\rho_{1}=-\frac{1}{3}(1-2\sqrt{2}iq_{a}) $  &\\ & $
\rho_{2,3}=\frac{1}{9}(-1 +i\sqrt{2}q_{a}\pm \sqrt{26}\sqrt{(1+4\sqrt{2}iq_{a})} $ &  no unmagnetized directions\\
\hline
$U_{9}^{q_{a}}U_{4}^{q_{b}})$ &$
\rho_{1}=1 \mbox{ , }
\rho_{2,3}=\pm i $& $
(1,0,0)$\\
\hline
$U_{9}^{q_{a}}U_{5}^{q_{b}})$ &
$\rho_{1}=-\frac{1}{3}(1+2\sqrt{2}iq_{a})  $  &\\ & $
\rho_{2,3}=\frac{1}{15}(-3-3\sqrt{2}iq_{a}\pm\sqrt{6}\sqrt{(11-22i\sqrt{2}q_{a})}$
&  no unmagnetized directions\\
\hline
$U_{9}^{q_{a}}U_{6}^{q_{b}}) $ &
$\rho_{1}=1  \mbox{ , }
\rho_{2,3}=\pm i$ &$
(1,0,0)$\\
\hline
\end{tabular}
\label{tab3}
\end{table}

\end{document}